\documentclass[aps,prl,twocolumn,showpacs,groupedaddress]{revtex4}  
\usepackage{color,graphicx,times}
\usepackage{amsmath,array}
\usepackage{multirow}
\usepackage{amssymb}
\usepackage{bm}

\newcommand{\ppb}{$p \bar p$}

\newcommand{\invpb}{pb$^{-1}$}
\newcommand{\invfb}{fb$^{-1}$}

\newcommand{\e}{e}%

\begin{document}

\hspace{5.2in} \mbox{FERMILAB-Pub-08-368-E}

\title{Search for large extra spatial dimensions
in the dielectron and diphoton channels \\
in $\bm{p\bar{p}}$ collisions at $\bm{\sqrt{s}=}$1.96\,TeV
}
 
%
\author{V.M.~Abazov$^{36}$}
\author{B.~Abbott$^{75}$}
\author{M.~Abolins$^{65}$}
\author{B.S.~Acharya$^{29}$}
\author{M.~Adams$^{51}$}
\author{T.~Adams$^{49}$}
\author{E.~Aguilo$^{6}$}
\author{M.~Ahsan$^{59}$}
\author{G.D.~Alexeev$^{36}$}
\author{G.~Alkhazov$^{40}$}
% ...

\affiliation{\vspace{0.1 in}(The D\O\ Collaboration)\vspace{0.1 in}}
% ...
\affiliation{$^{6}$University of Alberta, Edmonton, Alberta, Canada,
                Simon Fraser University, Burnaby, British Columbia, Canada,
                York University, Toronto, Ontario, Canada, and
                McGill University, Montreal, Quebec, Canada}
% ...
\affiliation{$^{29}$Tata Institute of Fundamental Research, Mumbai, India}
% ...
\affiliation{$^{36}$Joint Institute for Nuclear Research, Dubna, Russia}
% ...
\affiliation{$^{40}$Petersburg Nuclear Physics Institute,
                St. Petersburg, Russia}
% ...
\affiliation{$^{49}$Florida State University, Tallahassee, Florida 32306, USA}
% ...
\affiliation{$^{51}$University of Illinois at Chicago,
                Chicago, Illinois 60607, USA}
% ...
\affiliation{$^{59}$Kansas State University, Manhattan, Kansas 66506, USA}
% ...
\affiliation{$^{65}$Michigan State University,
                East Lansing, Michigan 48824, USA}
% ...
\affiliation{$^{75}$University of Oklahoma, Norman, Oklahoma 73019, USA}
% ...

\date{Sep 16, 2008}

\begin{abstract}
We report on a search for large extra spatial dimensions in the dielectron 
and diphoton channels using a  data sample of $1.05$ \invfb\ of \ppb\ collisions at a center-of-mass energy of 
1.96\, TeV collected by the D0 detector at the Fermilab Tevatron Collider. The invariant mass spectrum
of the 
data agrees well with the prediction of the standard model. We find 95\% C.L. lower limits on the effective Planck scale between 2.1 and 1.3 TeV for 2 to 7 extra dimensions.
\end{abstract}

\maketitle
Within the standard model (SM) the mass of Higgs boson is unstable against radiative corrections. The fact that the mass is not of the order of the GUT or Planck scales at $10^{16}$ or $10^{19}$~GeV but rather $\cal{O}$ ($10^2$~GeV) is commonly referred to as the ``hierarchy problem''. One way to circumvent the need for such fine tuning in the Higgs mass is by extending the dimentionality of the space, as in the large extra dimension model (LED) proposed by Arkani-Hamed, Dimopoulos and Dvali (ADD) \cite{ADD}, which posits that the fields of the standard model are pinned to a ($3+1$)-dimensional membrane, while gravity propagates in $n_{d}$ additional compactified spatial dimensions. Gauss' Law gives the relation between the fundamental Planck scale $M_{D}$, the observed Planck scale $M_{\text{Pl}}$, and the size of the extra dimensions $R$: $\big[M_{\text{Pl}}\big]^{2} \approx R^{n_{d}}\big[M_{D}\big]^{n_{d}+2}$. If $R$ is large compared to the Planck length $\simeq 1.6\times10^{-33}$~cm, $M_{D}$ can be as low as $\cal{O}$~(1~TeV), thus avoiding the hierarchy problem and making gravity strong at the TeV scale. Extra spatial dimensions will manifest themselves by the presence of a series of graviton states, known as a ``Kaluza-Klein tower", ($G_{KK}$). At colliders, large extra dimensions can be probed by searching for the effect of $G_{KK}$ on fermion or boson pair production \cite{ledexp}. 
\par
Extra dimension amplitudes will result in enhancement of the cross sections above the SM values, especially at high energies. The LED cross section, which consists of SM, interference, and direct gravity terms, can be parametrized by a single variable $\eta_{G}=F/M_{s}^{4}$ where $M_{s}$ is the the effective Planck scale, the ultraviolet cutoff of the sum over Kaluza-Klein states in virtual graviton exchange. The exact relationship between $M_{s}$ and $M_{D}$ depends on the exact quantum gravity scenario although they are of the same order of magnitude. The dimensionless parameter $F$ to leading order (LO) and the sub-leading $n_{d}$ dependence is given by
\begin{eqnarray}
 F&=& 1, \text{(GRW~\cite{grw})} \\
 F&=& \begin{cases} 
\ln(M_s^2/\hat {s}) \text{ for $n_{d}=2$}, \\
\frac{2}{n_{d}-2} \text{ for $n_{d}>2$ }
\end{cases}\text{(HLZ~\cite{hlz})}
\end{eqnarray}
where $\hat {s}$ is the center of mass energy of the partonic subprocess.
\par
In this Letter, we present a search for LED performed in events containing an $\e^+\e^-$ or $\gamma \gamma$ pair with $1.05$\,\invfb\ of $p
\bar{p}$ collider data collected with the upgraded D0\
detector \cite{det} between October $2002$ and February $2006$. With $127$ \,\invpb\ of data, D0\ has published limits on $M_{s}$ ranging from $0.97$ to $1.44$~TeV for $n_d = 7$ -- $2$ in the combined dielectron and diphoton channels \cite{59}. D0\ has also published limits in the dimuon channel with $246$ \,\invpb\ of data \cite{exp10}. The efficiency and resolution for high energy electromagnetic (EM) objects at D0\ are superior to those for muons and so a search for LED in combined $\e^+\e^-$ and $\gamma \gamma$ (di-EM) final states is superior to the dimuon channel. D0\ and CDF have also published limits on $M_{D}$ in the monophoton and monophoton plus monojet final states, respectively \cite{yuri}.
\par
Events are collected using triggers requiring the presence of at least one EM calorimeter shower with the transverse momentum with respect to the beam axis, $p_T$, greater than $15$~GeV. From these data we select $\e^+\e^-$ and $\gamma \gamma$ events using criteria that do not distinguish photons from electrons. We require events with two EM showers with $p_T > 25$ GeV. Showers are labelled CC (EC) if they are reconstructed in the central calorimeter (end cap calorimeters) with $|\eta| < 1.1$ ($1.5 < |\eta| < 2.4$), where pseudorapidity $\eta=-\ln[\tan(\theta/2)]$ and $\theta$ is the polar angle measured with respect to the proton beam direction. To reduce multijet background, we require at least one shower to be in the CC, so that selected events are either CC-CC (both showers in the CC) or CC-EC (one shower in the CC and the other in the EC). Each EM shower is required to be isolated, with less than $7\%$ of the cluster energy in an annular cone $0.2 < \Delta {\cal R} < 0.4$ about the shower centroid, where $\Delta {\cal R}=\sqrt {(\Delta\eta)^{2} + (\Delta\phi)^{2}}$ and $\phi$ is the azimuthal angle. We also demand the scalar sum of the $p_T$ of all tracks in the cone $0.05 < \Delta {\cal R} < 0.4$ be less than $2$ GeV. Finally, we demand the EM shower profile be consistent with that of an electron or photon using a $\chi^{2}$ test and that $97\%$ of the shower energy be contained in the EM calorimeter.
\par
The efficiencies for the electron and the photon selection criteria are determined from the same data set used for the event selection. We estimate separately the efficiencies for the $\chi^{2}$ requirement on the EM shower shape, the isolation requirements based on $\Delta {\cal R}$, and for all calorimeter-based high-$p_{T}$ triggers relevant to this analysis. In order to estimate the different efficiencies, we select a sample of di-EM events satisfying very loose EM identification requirements with invariant mass within $\pm 40$~GeV around Z boson mass. For each of these di-EM candidate events we estimate the efficiency as a function of $\eta$ versus $p_{T}$ using the tag and probe method ~\cite{tagprobe}. This efficiency is then applied to Monte Carlo simulation samples. 
\par
The irreducible background to the LED signal is from SM $\e^+\e^-$ and $\gamma \gamma$ production, while instrumental background arises from multijet and $\gamma$ + jet events with jets misidentified as EM objects. To model the invariant mass distribution of the physics backgrounds, we use the {\sc pythia}~\cite{pyth} event generator using the CTEQ6L1 parton distribution functions~\cite{pyth1}, followed by a {\sc geant}-based~\cite{pyth2}
detector simulation and reconstruction with the same algorithms as applied to data. The next-to-leading order (NLO) effect for both $\e^+\e^-$ and $\gamma \gamma$ is taken into account by multiplying the leading order (LO)
cross section by a mass independent $k$-factor of
1.34~\cite{Mathews:2004xp}. 
\par
We generate the LED signal for $2 \le n_d \le 7$ and $33$ different values of $M_s$ using a parton
level generator~\cite{standalone}. Following \cite{hlz}, we assume $Br(G_{KK}\rightarrow\gamma \gamma)/Br(G_{KK}\rightarrow\e^+\e^-)=2$. In order to model the effects of detector resolution and initial state radiation (ISR), we generate LED+SM and SM-only events separately to obtain the parton level distributions of the di-EM invariant mass versus the cosine of the scattering angle in the centre of mass frame of the two EM candidates ($|\cos \theta^{*}|$) for each value of $M_s$ and $n_{d}$ considered. The ratio of the LED+SM and SM 
distributions are obtained for all values of $M_s$ and $n_{d}$. Standard model events generated with the detailed {\sc geant}-based Monte Carlo simulation are weighted by this ratio to model the effect of an LED signal. We reweight the shape of the SM to simulate the LED signal, keeping the overall normalization as in the pure SM case. By normalizing to the Z boson production cross section (NNLO), where the signal contribution is negligible, we reduce the fractional uncertainty on the product of the efficiency and integrated luminosity.
\par
\begin{table*}[!t]
\caption{\label{baycc} Number of events observed and expected from SM processes in different mass windows for CC-CC and CC-EC events. The individual contributions to the total SM expectation from multijet, $\e^+\e^-$ and $\gamma \gamma$ are also shown separately. }
\begin{ruledtabular}
\begin{tabular}{ccrr@{\ $\pm$\!\!\!\!}lr@{\ $\pm$\!\!\!\!}lllcrr@{\ $\pm$\!\!\!\!}lr@{\ $\pm$\!\!\!\!}lll }
& & \multicolumn{7}{c}{CC-CC} & & \multicolumn{7}{c}{CC-EC} \\ 
\cline{1-1} \cline{3-9} \cline{11-17} 
 Mass & & \!\!\! Data & \multicolumn{2}{c}{\!\!\!\!\!\!\!Total Background} & \multicolumn{2}{c}{\!\!\!\!\!\!Multijet}(MJ)& $\;\e^+\e^-$ & $\!\!\gamma \gamma$ & & Data & \multicolumn{2}{c}{\!\!\!\!\!\!\!Total Background} & \multicolumn{2}{c}{\!\!\!\!\!\!Multijet}(MJ)& $\;\e^+\e^-$ & \!\!$\gamma \gamma$\\    
(GeV)        & & $N$              & \ $N_{b}$ & $N_{b}^{\text{sys}}$  & \!\!\!\!$N_{\text{MJ}}$ & $N_{\text{MJ}}^{\text{sys}}$ & $N_{\e^+\e^-}$ & \!\!$N_{\gamma \gamma}$ & & $N$ & \;$N_{b}$ & $N_{b}^{\text{sys}}$  & $N_{\text{MJ}}$ & $N_{\text{MJ}}^{\text{sys}}$ & $N_{\e^+\e^-}$ & \!\!$N_{\gamma \gamma}$ \\
\cline{1-1} \cline{3-9} \cline{11-17} 

240--290  & & 61 &\, 67    & 8    & 22     & 3.1  &\; 30   & \!\!15 & & 144 &\!\! 171    & 34    & 115    & 34    & 34    & \!\!21\\
290--340  & & 30 &\, 28    & 4    &  7     & 1  &\; 14   & \!\!7 & &  52 &\!\!  55    & 11    &  35    & 11    & 12    &  \!\!8\\
340--400  & & 21 &\, 15    & 2    &  3     & 1  &\; 7    & \!\!5 & &  21 &\!\!  23    &  5    &  12    &  4    &  7    & \!\!4 \\
400--500  & &  9 &\,  9    & 1  &  1.4   & 0.3  &\; 5    & \!\!3 & &  12 &\!\!   9    &  2    &   4    &  2  &  3.3  &  \!\!1.2\\
500--600  & &  1 &\, 3.6    & 1.2  &  0.14  & 0.09 &\; 2.4  & \!\!1.1 & &   2 &\!\!   1.5    &  0.4 &   0.6 &  0.2 &  0.73 &  \!\!0.18\\
600--1000 & &  2 &\,  1.3  & 0.1 &  0.11  & 0.06 &\; 0.67 & \!\!0.53 & &   0 &\!\!   0.35 &  0.07  &   0.03 &  0.04 & 0.24 &  \!\!0.08 \\
\end{tabular}
\end{ruledtabular}
\vspace{-2mm}
\end{table*}
\begin{figure*}[!t]
\begin{tabular}{cc}
\includegraphics[scale=0.4]{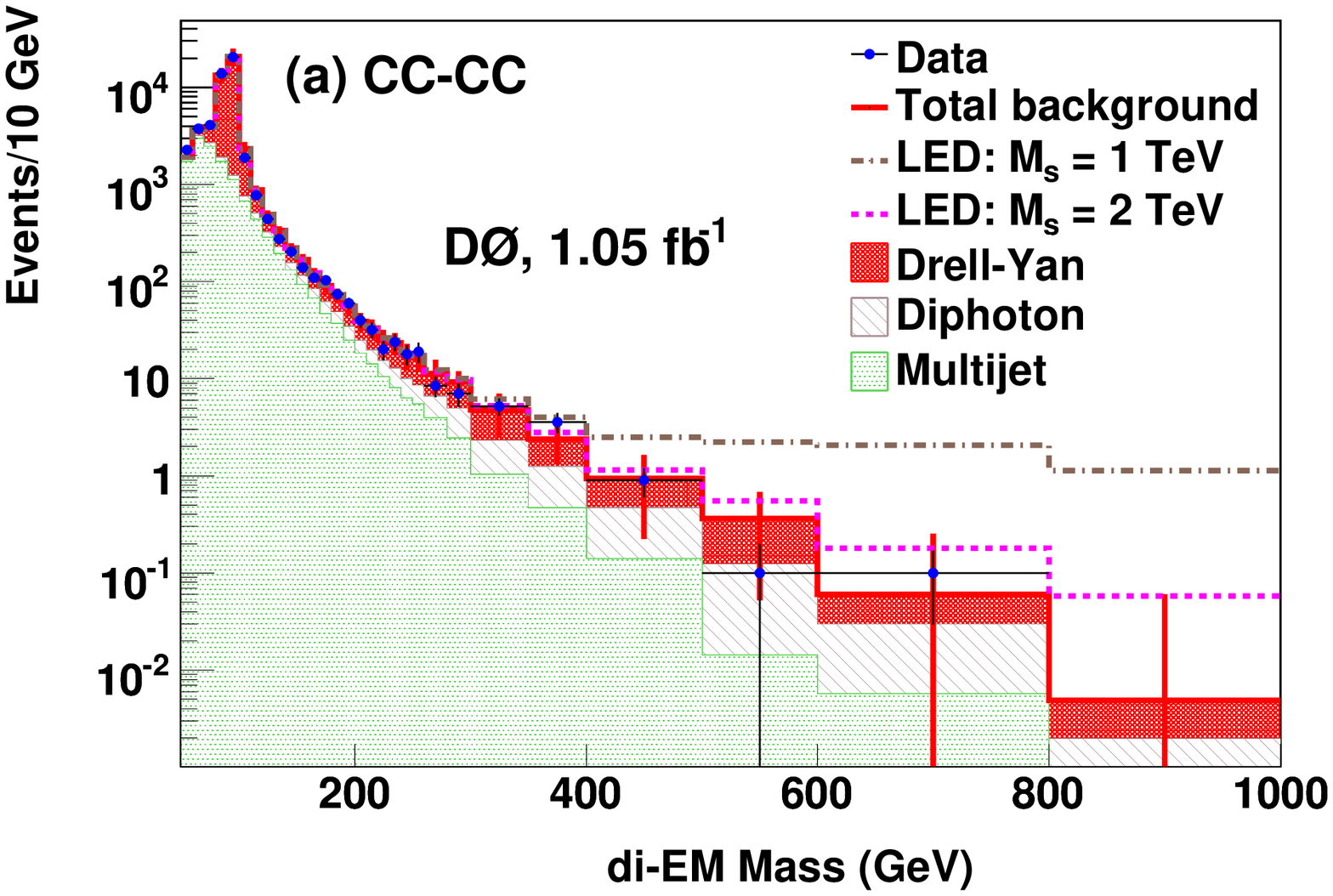} &
\includegraphics[scale=0.4]{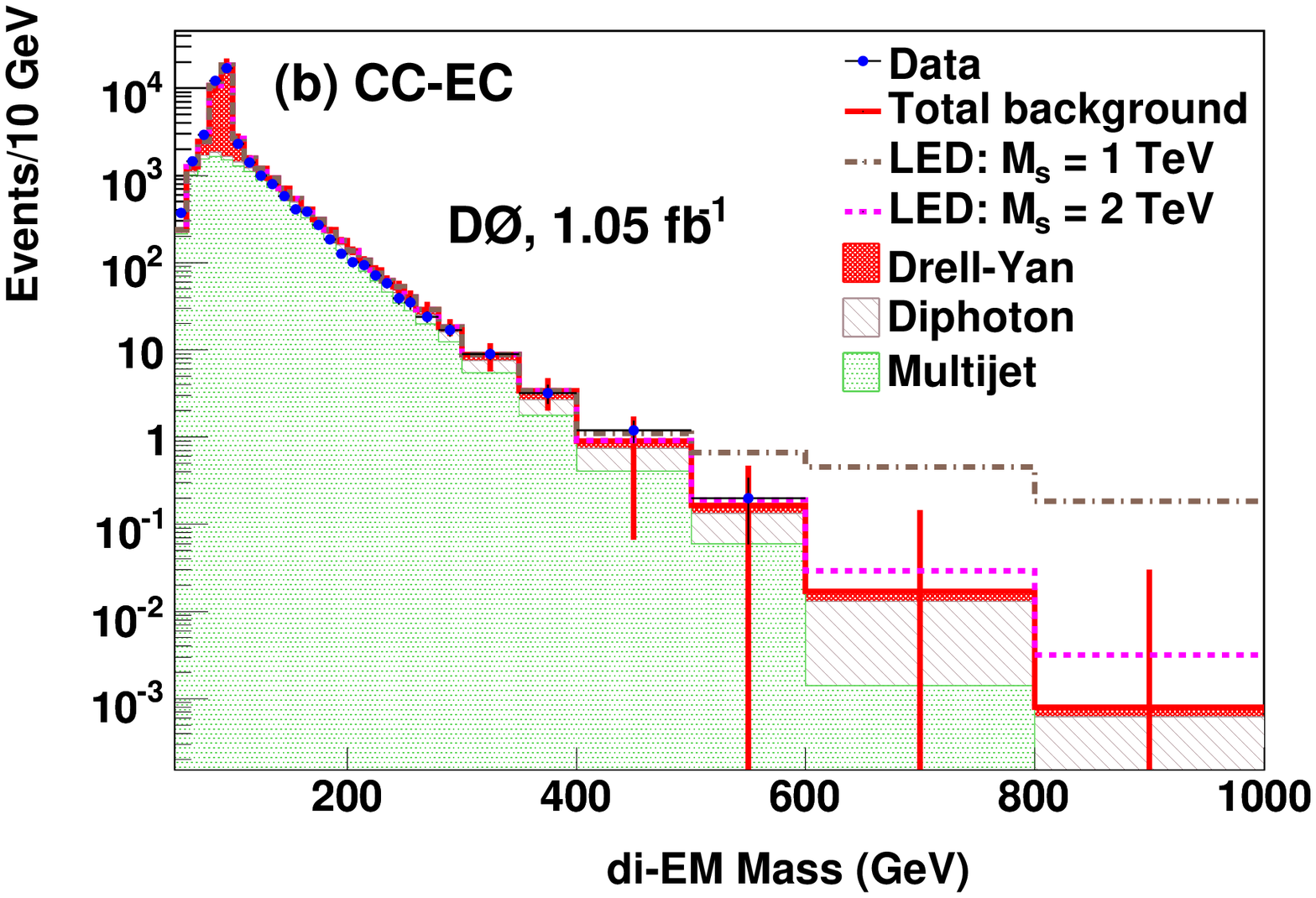}
\end{tabular}
\vspace{-2mm}
\caption{\label{cccc-4}The di-EM invariant mass distributions for CC-CC (a) and CC-EC (b) 
events. The data are shown by points with error bars, the filled histograms represent the Drell-Yan, diphoton and multijet backgrounds, and the solid line represents the total background.
The broken lines show the invariant mass distributions for two different values of $M_{s}$ for $n_{d}=4$. The error bars for the total background include both statistical and systematic uncertainties.}
\vspace{-0.3cm}
\end{figure*}
\begin{figure*}[!t]
\begin{tabular}{cc}
\includegraphics[scale=0.4]{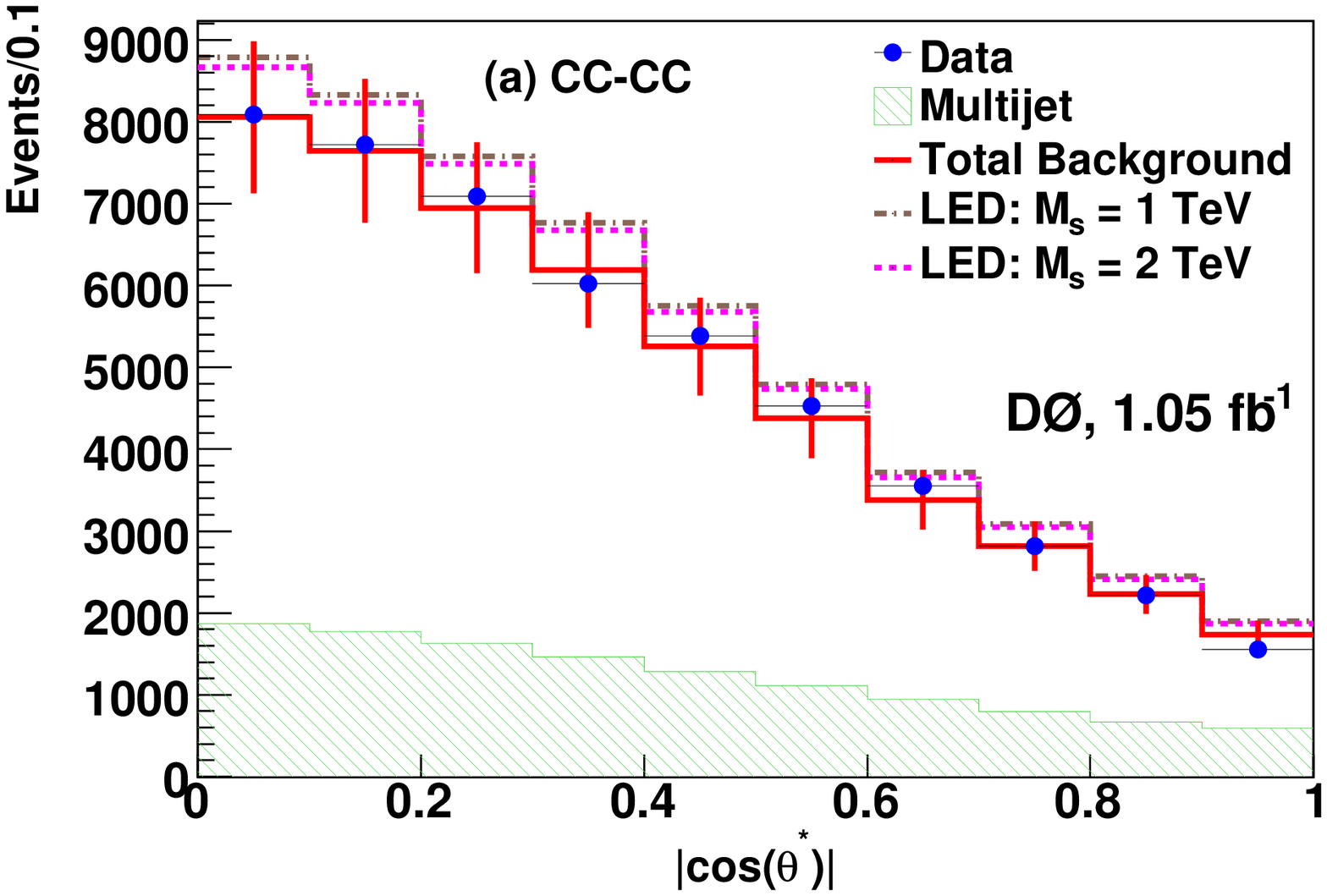} &
\includegraphics[scale=0.4]{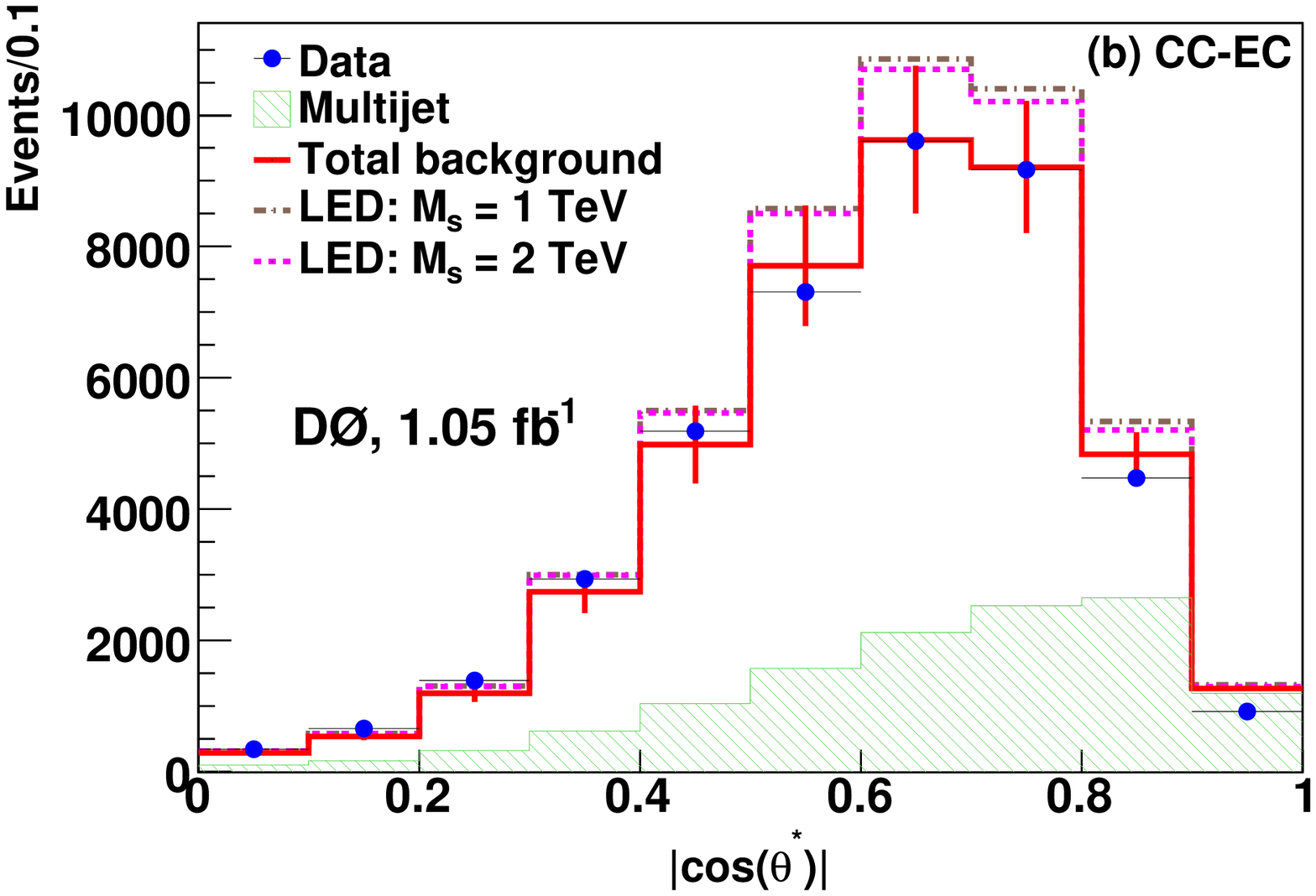}
\end{tabular}
\vspace{-2mm}
\caption{\label{coscccc-4}The distributions of the center-of-mass scattering angle $\cos \theta^{*}$ of the two final state EM candidates in CC-CC (a) and CC-EC (b) events. The data are shown by points with error bars, the filled histogram represent the multijet background, and the solid line represents the total background. The broken lines show the $\cos \theta^{*}$ distributions for two different values of $M_{s}$ for $n_{d}=4$. The error bars for the total background include both statistical and systematic uncertainties.}
\vspace{-0.4cm}
\end{figure*}
To estimate the normalization of the multijet background, we fit the di-EM invariant mass distribution of the selected data events with a 
linear combination of the physics and instrumental background distributions. The shape of the invariant mass
distribution for the instrumental background is estimated from data events
with EM energy clusters that
fail the $\chi^{2}$ requirement for the shower profile. This fit is performed in the mass range $60$ -- $140$ GeV where we
expect no contribution from LED. We obtain separate fits for CC-CC and CC-EC events. From the fits we determine the fraction $f_{\rm MJ}$ of the multijet contribution to the total background in the mass range $60$ -- $140$ GeV to be
$f_{\rm MJ}=0.22 \pm 0.03$ in CC-CC events and $f_{\rm MJ}=0.24 \pm 0.07$ in
CC-EC events. We extrapolate the
total background using the fitted value of $f_{\rm MJ}$ to determine the expected number of background
events with invariant mass above $140$ GeV in both the CC-CC and
CC-EC configurations. Table~\ref{baycc} 
shows the numbers of events in different mass ranges for CC-CC and CC-EC where we would expect the LED signal to appear. The number of events is consistent with the number of
expected events from the SM expectation. Figure\ \ref{cccc-4}(a) shows the invariant mass
distribution for CC-CC events and Fig.\ \ref{cccc-4}(b) for CC-EC events. The distributions of $|\cos
\theta^{*}|$ are shown in Fig.\ \ref{coscccc-4} for CC-CC and CC-EC both for data and the background model. We
find that the total background distribution for the invariant mass and $|\cos \theta^{*}|$
is consistent with the data within statistical and systematic
uncertainties. 
\par
Most of the systematic uncertainties on the background model are dependent on the invariant mass. The dominant uncertainty arises from the efficiency of the $\chi^{2}$ cut on the shower profile used to estimate multijet background (13\% of the background itself in CC-CC and 30\% in CC-EC). The systematic on the LED modeling is dominated by uncertainties on the choice of parton distribution functions \big[(1--19)\% in CC-CC and (1.5--12)\% in CC-EC\big]. All the other signal uncertainties are correlated to SM background uncertainties due to the technique used to generate our LED signal. Table~\ref{syst} summarizes the dominant background and signal uncertainties taken into account in calculating the limit on $M_{s}$. The NLO k-factor uncertainty refers to the uncertainty due to choice of PDF, renormalization and factorization scale.
\begin{table}[!phtb]
\vspace{-5mm}
\caption{\label{syst} Systematic uncertainties (in \%) on the predicted numbers of signal and background events considered in calculating the limit on $M_{s}$.}
\begin{ruledtabular}
\begin{tabular}{llccc}
   &    &                      CC-CC   &        &    CC-EC  \\
\hline
Signal only    &   &                   &        &           \\
   & Acceptance  &        1--19   &        &   1.5--12 \\
   & Luminosity     &                  &    4   &           \\
\hline
Signal and            & &              &        &           \\
background            & &              &        &           \\
   & Trigger + EM selection  &  6      &        &     5     \\
   & Energy scale     &        5--13   &        & 0.3--3.5  \\
   & Energy resolution  &    0.3--1.7  &        & 0.2--3.5  \\
   & NLO k-factor   &                  & 3--10 &           \\
   & k-factor mass dependence      &                  &   5    &           \\	
   & PDF            &                  & 5.5--9 &           \\		
\hline
Background only  & &                   &        &           \\
   & Multijet    &             13     &        &   30      \\	
\end{tabular}
\end{ruledtabular}
\end{table}
\vspace{-0.1cm}
\par
The two-dimensional distribution of the invariant mass and $|\cos \theta^{*}|$ for the observed dielectron and diphoton
events is compared with the corresponding
distributions expected from SM physics and instrumental background,
and the LED signal for $M_s$ ranging from $1$~TeV to $3$~TeV for a given $n_{d}$. The posterior probability
density $P(M_{s} \mid \text{Data})$ given the number of observed events in the $k^{th}$ mass bin and $l^{th}$ $\cos \theta^{*}$
bin, $N^{k,l}_{\text{obs}}$, is then computed using a Gaussian prior for the SM plus multijet background. Evidence of large extra dimensions with a given $M_{s}$ will
appear as a peak in $P(M_{s} \mid \text{Data})$ distribution. In the absence of signal we proceed to estimate the lower limit on $M_{s}$ using the semi-frequentist CLs method \cite{coll}, which is based on computation
of a log likelihood ratio. Both the expected and observed limits on $M_s$ at the 95\%~C.L. are calculated. Systematic uncertainties in the signal and background distributions
are taken into account in this calculation, with their correlations properly included. The distribution of the ratio of the observed (expected) upper limit at the 95\%~C.L. limit to the predicted cross section as a function of $M_{s}$ is used to extract the observed (expected) limit on $M_s$ for $n_{d}=7$ to $n_{d}=2$.
\par
\begin{figure}[!t]
\includegraphics[scale=0.4]{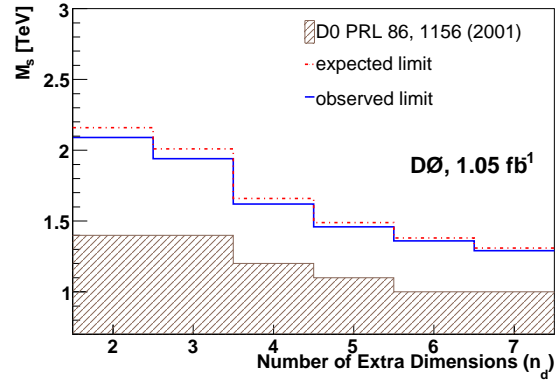} 
\vspace{-2mm}
\caption{\label{cumu4} Observed and expected limits on the effective Planck scale, $M_{s}$, in the di-EM channel along with previously published limits in di-EM channel.}
\vspace{-6mm}
\end{figure}

For the $n_{d}$ independent GRW formalism, we calculate the observed(expected) limit on $M_{s}$ of $1.62(1.66)$~TeV. We obtain the observed limits on $M_s$ at the 95$\%$~C.L. in the HLZ formalism (sub-leading, $n_{d}$ dependent) ranging from $1.29$ to $2.09$~TeV for $n_{d}=7$ to $n_{d}=2$. Both the observed and expected limits on $M_s$, for different formalisms and for six different $n_{d}$ are summarized in Table~\ref{limit}. The observed and expected limits on $M_{s}$ for a given number of extra dimensions are found to be similar. The present limits are a significant improvement over the published limit \cite{59}. Figure\ \ref{cumu4} summarizes the observed and expected limits on $M_{s}$ along with the previously published limits on $M_{s}$ in the di-EM channel.
\begin{table}[!t]
\caption{\label{limit} Observed and expected lower limits at the 95\%~C.L. on the effective Planck 
scale, $M_{s}$, in TeV.}
\begin{ruledtabular}
\begin{tabular}{ccccccccccc}
     & & GRW  & &       & \multicolumn{6}{c }{HLZ} \\ 
\cline{3-3} \cline{5-11}
     & &      & & $n_d$ & 2 & 3 & 4 & 5 & 6 & 7 \\ 
Obs. & & 1.62 & &       & 2.09 & 1.94 & 1.62 & 1.46 & 1.36 & 1.29 \\
Exp. & & 1.66 & &       & 2.16 & 2.01 & 1.66 & 1.49 & 1.38 & 1.31 \\
\end{tabular}
\end{ruledtabular}
\vspace{-4mm}
\end{table}
\par
In summary, we have performed a dedicated search for large extra spatial dimensions by looking for effects of virtual Kaluza-Klein graviton in the dielectron and diphoton channels using $1.05$ \invfb\ of data collected by D0\ detector. We see no evidence of excess over the standard model prediction and set limits at 95\%~C.L. on the effective Planck scale at $2.09$(1.29)~TeV for $2$(7) extra dimensions. These are presently the most restrictive limits on large extra dimensions. 

%
We thank the staffs at Fermilab and collaborating institutions, 
and acknowledge support from the 
DOE and NSF (USA);
CEA and CNRS/IN2P3 (France);
FASI, Rosatom and RFBR (Russia);
CNPq, FAPERJ, FAPESP and FUNDUNESP (Brazil);
DAE and DST (India);
Colciencias (Colombia);
CONACyT (Mexico);
KRF and KOSEF (Korea);
CONICET and UBACyT (Argentina);
FOM (The Netherlands);
STFC (United Kingdom);
MSMT and GACR (Czech Republic);
CRC Program, CFI, NSERC and WestGrid Project (Canada);
BMBF and DFG (Germany);
SFI (Ireland);
The Swedish Research Council (Sweden);
CAS and CNSF (China);
and the
Alexander von Humboldt Foundation (Germany).
%

\end{document}